\def\aap{\mbox{\bf{Astron.\ Astrophys.}}}

\def\apjl{\mbox{\bf{Astrophys.\ J.\ L.}}}

\def\apjs{\mbox{\bf{Astrophys.\ J.\ Suppl.}}}
\documentclass[letter, traditabstract]{aa} 
\usepackage{amsmath}
\usepackage{longtable, lscape}
\usepackage{graphicx}
\usepackage{txfonts}
\usepackage{natbib}
\usepackage{color}

\bibpunct{(}{)}{;}{a}{}{,} 

\newcommand{\Msun}{\mbox{$~\mathrm{M}_{\odot}$}}

\begin{document}
%

   \title{Massive binaries as the source of abundance anomalies in globular clusters}

   \author{S.E. de Mink\inst{1} \and O.R. Pols \inst{1} \and
   N. Langer\inst{2,1} \and R.G. Izzard \inst{3} }
   \institute{Astronomical Institute, Utrecht University, PO Box
   80000, 3508 TA Utrecht, The Netherlands \and Argelander-Institut
   f\"{u}r Astronomie der Universit\"{a}t Bonn, Auf dem H\"{u}gel 71,
   53121 Bonn, Germany \and Universit\'e Libre de Bruxelles, Boulevard
   du Triomphe, B-1050 Brussels, Belgium \\ S.E.deMink@uu.nl;
   O.R.Pols@uu.nl; nlanger@astro.uni-bonn.de; Robert.Izzard@ulb.ac.be;
   }
   \date{Received 29/08/2009; accepted 2/10/2009 }

 
   \abstract {Abundance anomalies observed in globular cluster stars
     indicate pollution with material processed by hydrogen
     burning. Two main sources have been suggested: asymptotic giant
     branch (AGB) stars and massive stars rotating near the break-up
     limit (spin stars).
     We propose massive binaries as an alternative source of processed material.
     We compute the evolution of a 20\Msun~star in a close binary
     considering the effects of non conservative mass and
     angular momentum transfer and of rotation and tidal interaction to
     demonstrate the principle. We find that this system sheds about
     10\Msun~of material, nearly the entire envelope of the primary
     star. The ejecta are enriched in  He, N, Na, and Al and depleted
     in C and O, similar to the abundance patterns observed in gobular
     cluster stars. However, Mg is not significantly depleted in the ejecta of this model.

     In contrast to the fast, radiatively driven winds
     of massive stars, this material is typically ejected with low
     velocity. We expect that it remains inside the potential well of
     a globular cluster and becomes available for the formation or
     pollution of a second generation of stars.
     We estimate that the amount of processed low-velocity material
     ejected by massive binaries is greater than the contribution of AGB
     stars and spin stars combined, assuming that the majority of
     massive stars in a proto-globular cluster interact with a
     companion and return their envelope to the interstellar medium.
     If we take the possible contribution of intermediate mass stars in binaries into account and assume that the ejecta are diluted  
     with an equal amount of unprocessed
     material, we find that this scenario can potentially provide enough
     material to form a second generation of low-mass stars, which is
     as numerous as the first generation of low-mass stars,
     without the need to make commonly adopted assumptions, such as
     preferential loss of the first generation of stars, external
     pollution of the cluster, or an anomalous initial mass function.

}


   \maketitle
%

\section{Introduction}

For a long time star clusters have been considered as idealized
single-age, chemically homogeneous stellar populations. However, it
has recently become clear that many clusters show multiple main
sequences and sub giant branches and extended horizontal branches
\citep[e.g.][]{Piotto+07}, implying the existence of multiple
populations within one cluster.%
%
%
In addition, large star-to-star abundance variations are found for
light elements such as C, N, O, Na, and Al, while the composition of
heavier elements (Fe-group and $\alpha$-elements) seems to be
constant.  Field stars with the same metallicity do not exhibit these
abundance patterns \citep[for a review see][]{Gratton+04}.
These chemical variations have been interpreted as originating from the presence of both a ``normal'' stellar population, exhibiting abundances similar to field stars of the same metallicity and a second  population of stars formed out of material processed by hydrogen burning via the CNO-cycle and by the NeNa and MgAl chains \citep[e.g.][]{Prantzos+07}.  {According to \citet{Carretta+09},  50-70\% of the stars in gloular clusters belong to the second population.}

Two sources of processed ejecta have been proposed: the slow winds
  of \emph{massive AGB stars}, which enrich their convective envelopes
  with H-burning products
  \citep{Ventura+01,Dantona+02,Denissenkov+Herwig03} and fast-rotating
  massive stars (we refer to these as \emph{spin stars}), which
  are believed to expel processed material centrifugally when they
  reach break-up rotation
\citep{Prantzos+Charbonnel06,Decressin+07a}.%
%
%
In this scenario a first generation of stars is formed out of pristine
material. Their low-velocity ejecta are trapped inside the potential
well of the cluster and provide the material for the formation of a
second generation of stars.
%
%
%
Although both proposed sources are promising, matching the observed
abundance patterns and providing enough ejecta for the formation of a
second generation that outnumbers the first generation have proven to
be two major challenges.
In this Letter we propose \emph{massive binaries} as a candidate
for the internal pollution of globular clusters.  

\begin{figure*}
\centering
\includegraphics[angle=-90, width=1.0\textwidth, bb=50 50 283 616]{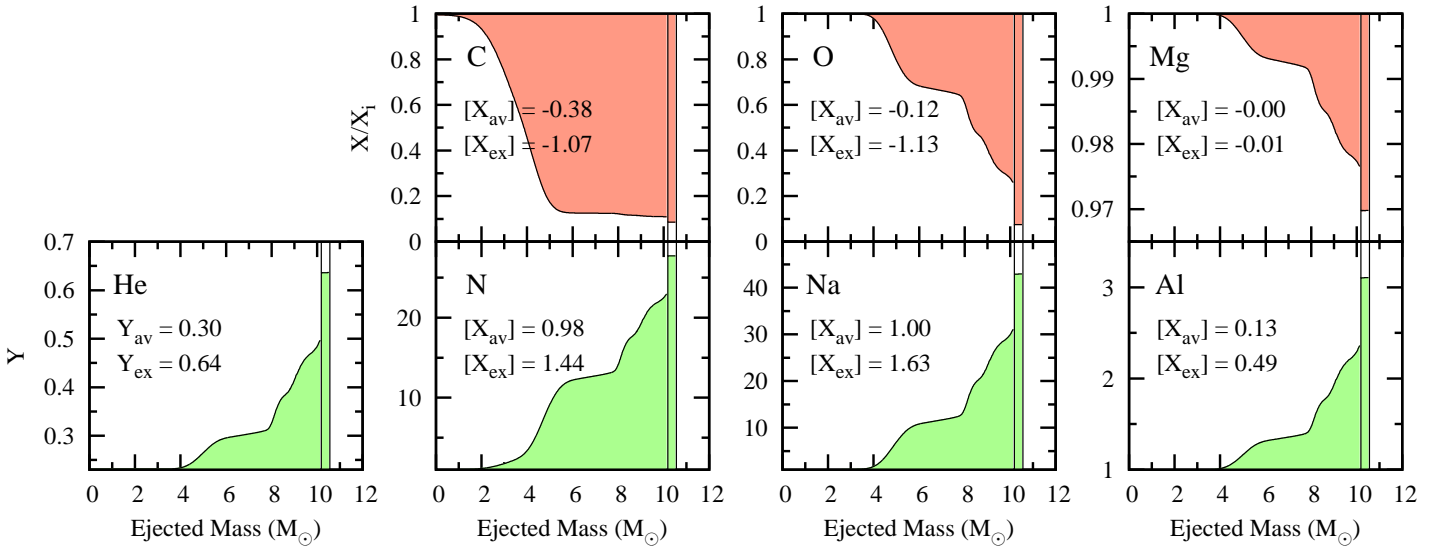}
\caption{ Composition of the slow ejecta of the modeled binary system 
(Sec.~3) as a function of the ejected amount of mass.  The mass
  fraction $X$ of the main stable isotope of each element is given
  relative to the initial mass fraction $X_{\rm i}$, except for Mg
  where we added $^{24}$Mg$,^{25}$Mg and $^{26}$Mg.  The average
  $X_{\rm av}$ and the most extreme mass fraction $X_{\rm ex}$ are
  given in each panel on a logarithmic scale: $[X] \equiv \log_{10} (X
  / X_{\rm i})$. For helium we show the absolute mass fraction $Y$
  instead. Mass ejected during the first and second mass transfer
  phases is separated by a thin vertical line. \label{fig:chem}}
\end{figure*}

\section{Binaries as sources of  enrichment}

Interacting binaries can shed large amounts of material processed by
hydrogen burning into their surroundings.  A clear example is the
well-studied system RY Scuti. It is undergoing rapid mass transfer
from a 7\Msun~supergiant to its 30\Msun~companion. Mass is lost from
the system via the outer Lagrangian points into a circumbinary disk
and a wider double toroidal nebula.  The nebula shows signatures of
CNO processing: it is enriched in helium and nitrogen and depleted in
oxygen and carbon
\citep{Smith+02,Grundstrom+07}. In contrast to the high-velocity
radiatively driven winds of massive stars, these ejecta have low
velocities. \citet{Smith+01} measure expansion velocities ranging from
30 to 70 km~s$^{-1}$ in the nebula of RY Scuti, which are lower than
the present-day escape velocity of massive globular
clusters. Furthermore, the nebula shows evidence of clumps
\citep{Smith+02} and dust in the outer parts \citep{Gehrz+01}, which
may serve as seeds for forming a second generation of
low-mass proto stars.

Evidence of severe mass loss from interacting binaries comes
from a wide variety of observed interacting and post-interaction
systems. This appears to be a common phenomenon for many interacting
binaries.
Various authors have inferred highly non conservative evolution for
Algols, systems that are currently undergoing stable mass transfer
\citep[e.g.][]{Refsdal+74,deGreve+Linnell94, Figueiredo+94,
VanRensbergen+06, DeMink+07}. Most notable are short-period
 binaries containing a compact object, e.g. cataclysmic variables,
 X-ray binaries, binary radio pulsars and double white dwarf
 systems. Their formation requires a phase of severe mass and angular
 momentum loss by ejection of a common envelope. Direct evidence of
 this type of evolution comes from planetary nebulae with close binary
 nuclei, which appear to have recently emerged from the
 common-envelope phase \citep[for a review see][]{Iben+Livio93}.

Theoretical considerations support the idea that most interacting
binaries shed large amounts of mass. Three-dimensional hydrodynamical
simulations of the mass transfer stream and accretion disk of the
interacting binary $\beta$ Lyrae predict that 50\% of the transferred
mass is lost \citep{Bisikalo+00, Nazarenko+Glazunova06A}. In addition,
\citet{Ulrich+Burger76} showed that the accreting star is driven out of
thermal equilibrium and expands. This can lead to contact and to strong
mass and angular momentum loss from the system
\citep{Flannery+Ulrich77}.
Furthermore, \citet{Packet81} noted that the accreting star reaches
break-up rotation after gaining only a few percent of its own mass.
Rapid rotation is found for many accreting stars in Algols
\citep{Barai+04} and this mechanism has been proposed to explain the
formation of Be-X-ray binaries \citep[e.g.][]{Pols+91}. In principle,
tides can counteract the effect of spin-up by mass transfer in close
binaries. \citet{Petrovic+05_WR} have computed detailed binary evolution
models taking these effects into account. They find that massive
binaries with initial periods as short as 3-6 days lose 70-80\% of the
transferred mass on average. For wider and more massive systems, they
expect even less conservative mass transfer, such that nearly the
entire envelope of the primary is returned to the interstellar medium.

%


\section{Composition of the ejecta\label{sec:model}} 
To investigate the yields of a typical massive binary, we employ a
state-of-the-art binary evolution code described by
\citet{Petrovic+05_GRB} and \citet{Yoon+06}. The effect of mass and
angular momentum loss on the binary orbit is computed according to
\citet{Podsiadlowski+92}, with the specific angular momentum of the
wind calculated according to \citet{Brookshaw+93}. Mass transfer is
modeled following \citet{Ritter88}. Tidal interaction is modeled as
described in \citet{Detmers+08}.  Non conservative mass transfer is
modeled self-consistently: it results from the interplay between
spin-up by mass transfer, tidal interaction, and rotationally enhanced
mass loss. To follow the nucleosynthesis up to the advanced stages of
hydrogen burning, we updated the reaction rates to the NACRE 99
compilation \citep{Angulo+99}.  As initial composition we assumed an
$\alpha$-enhanced mixture with a metallicity of $Z=5 \times 10^{-4}$,
following \citet{Decressin+07a}.  
We assumed masses of 20 and 15\Msun~for the two stars, an orbital
period of 12 days and initial rotation rates synchronized with the
orbital revolution. Because of these low initial rotation rates, the
effect of rotationally induced mixing on the low-velocity ejecta is
negligible. We follow the evolution from the onset of
hydrogen burning until central carbon burning.

Shortly after hydrogen exhaustion in the center, the primary expands and
starts to transfer mass to its companion. Initially the secondary star
efficiently accretes all the transferred mass and associated
angular momentum and spins up.  After accreting about 1.5\Msun~it
approaches critical rotation. From this moment on, the majority of
the transferred mass is ejected from the system.

The transition from conservative to non conservative mass transfer,
when the accreting star reaches critical rotation, provides an
interesting selection mechanism.  Initially, when the outermost
unprocessed layers are transferred, the companion star efficiently
accretes all the material. By the time deeper layers of the donor star,
which do show signatures of nuclear processing, are exposed, mass is
 ejected from the system.

After transferring nearly its entire envelope, the donor contracts,
ignites helium, and becomes a Wolf-Rayet star.  In the meantime the
secondary spins down because of angular momentum loss in its
rotationally enhanced wind. A second phase of mass transfer sets in
when the Wolf-Rayet star expands during He-shell burning. This
time about 1\Msun~ is transferred, which is initially accreted by the
secondary but ejected shortly afterwards by its rotationally enhanced
wind.
Our computation ends after the primary star ignites carbon and fills
its Roche lobe a third time. Soon after, it will explode as a Type~Ib/c
supernova.

Figure~\ref{fig:chem} shows the composition of the material
ejected from the system during the two phases of mass transfer. The
first 2\Msun~are relatively unprocessed and resemble the pristine
composition, except for the depletion of fragile elements such as
lithium (not plotted). The next 2\Msun~of ejected material are  processed by
CN cycling (nitrogen being enhanced up to a factor five), followed by
about 4\Msun~showing He enrichment and the O-Na
anti-correlation. After the ejection of 8\Msun, a sudden change in
slope is visible for all elements except carbon. 
The layers of the donor star, which are exposed at this moment, were
part of the convection zone above the H burning shell. Here, the
temperatures were high enough for proton captures onto $^{25}$Mg and
$^{26}$Mg, leading to an increase in the aluminum abundance by a factor
two. Between the first and second mass transfer phases, the primary star
loses mass in the form of a fast Wolf-Rayet wind, and even deeper
layers of the primary are exposed.  Because the these high-velocity winds are likely to escape from the cluster, we have exclude the mass ejected during this  phase in Fig.~\ref{fig:chem}.


The average composition of the ejecta are comparable to the yields of a 60\Msun~spin star model computed by \citet{Decressin+07a} by adopting the same set of reaction rates. For example, we find an average helium mass fraction of 0.30 and an enhancement in sodium by 1.0 dex compared to 0.32 and 1.3 dex, respectively, in the 60\Msun~spin star. The sum of carbon, nitrogen, and oxygen is constant within a few percent, consistent with observations \citep[see references in][]{Decressin+09}.
However, the temperatures are not high enough for efficient proton
captures onto the most abundant isotope of magnesium, $^{24}$Mg,
adopting the recommended reaction rates. The onset of this reaction is
required to explain the full range of aluminum abundances observed in
some globular clusters.  We expect that considering more massive or
wider binaries (in which mass transfer starts in a later evolution
stage) might alleviate this problem.

In shorter-period binaries, tides can counteract the effect of spin-up:
more mass is accreted before the accreting star reaches break-up
rotation and less mass is lost from the system. However, the ejecta
originate from the last exposed (deepest) layers of the donor
star and will therefore show more strongly pronounced
anti-correlations.
For initially wider systems, we expect that nearly the complete
envelope is ejected. In addition stars may have considerable rotation
rates at birth. For these stars, mixing induced by rotation can lead to
processing of the whole envelope {\citep[e.g.][]{Yoon+06, Decressin+07a}}.

Besides the effects discussed above, the (remainders of the) binary can still shed large amounts of H-processed material at low velocities in various ways. The companion star is now rapidly rotating and may reach break-up rotation again towards the end of its evolution as proposed in the spin star scenario. Processed mass can be ejected during a phase of reverse mass transfer from the secondary onto the compact object, if the system remains bound after the supernova explosion of the primary star.

\begin{figure}
\centering
\includegraphics[width=0.5\textwidth]{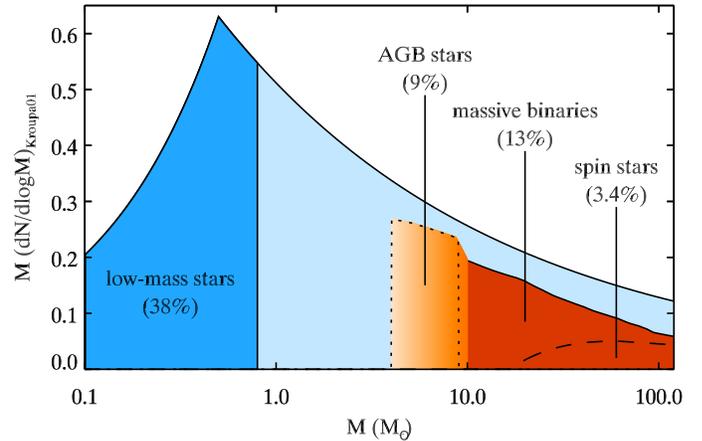}
\caption{ 
Mass-weighted \citet{Kroupa01} IMF as a function of stellar mass. The
 surface areas indicate the mass contained in the first generation of
 long-lived low-mass stars (dark blue), the ejecta of AGB stars
 (dotted line), spin stars, i.e. fast-rotating massive stars (dashed line), and massive (red) and
 intermediate-mass (orange) binaries. Percentages indicate the
 fraction of mass relative to the total mass contained in stars of the
 first generation. See Sect.~\ref{sec:pop} for
 details. \label{fig:pop}}
\end{figure}

\section{Mass budget\label{sec:pop}}

One of the main challenges for the two previously proposed sources of
pollution, massive AGB and spin stars, is to provide the large amount
of ejecta needed to create a second population that is larger than
the first population.
The population of low-mass stars (0.1-0.8\Msun), which can still be observed today, represents 38\% of the stellar mass initially present in the cluster assuming a standard \citet{Kroupa01} initial mass function (IMF) between 0.1-120\Msun, see Fig~\ref{fig:pop}.  The ejecta of AGB stars with initial masses between 4 and 9\Msun~represent up to 8.9\% of the initial stellar mass \citep[assuming an initial-final mass relation by][]{Ciotti+91}.  For spin stars this fraction is 3.4\%, if one assumes that every massive star is single and born with a rotational velocity high enough to reach break-up rotation \citep[using models by][]{Decressin+07a}. There are not enough of these ejecta to create a second generation which is equally numerous as the first generation, even when we assume that the second generation consists only of low-mass stars and that star formation is very efficient, see Fig~\ref{fig:pop}.

Two rather extreme solutions have been proposed.
(1) The IMF was highly anomalous, favoring the formation of the
polluting stars with respect to the long-lived low-mass stars that we
observe today. Even though we have no direct constraints on the IMF of
globular clusters,
\citet{Kroupa02} finds that the IMF is remarkably uniform in stellar
populations with very different properties.
(2) Clusters were initially at least 10-20 times more massive and they
have primarily lost low-mass stars from the first generation as a
result of the dynamical evolution and tidal stripping of the cluster
\citep{Decressin+07b, Decressin+08, D'Ercole+08}.
In this section we investigate to what extent the ejecta of massive
binaries can alleviate this conundrum.

Even though the current fraction of detected binaries in globular
clusters is not high \citep[e.g.][]{Davies+08}, this is not
necessarily the case for the high-mass stars originally present in the
cluster. \citet{Sana+08} and \citet{Mason+09} find a minimum binary
fraction of 60-75\% for O stars associated with clusters or OB
associations.  In globular clusters, these fractions may even be
higher.  In this environment, close binaries can be created and
tightened during and after the star formation process, for example by
the dissipative interaction with gas \citep[e.g.][]{Bonnel+Bate05} and
by three-body interactions, such as the Kozai mechanism in combination
with tidal friction \citep{Fabrycky+07}.  With massive stars
preferentially residing in the dense core of the cluster, where the
dynamical encounters are most frequent, it is not unreasonable to
assume that the large majority of massive stars interact by mass
exchange.

Let us assume that every massive star is a member of an interacting
binary.  In Sects.~2~and~3, we argued that nearly the entire envelope
of the donor is returned to the ISM.  For simplicity, we neglect the
contribution of the secondary star after it has been spun up by mass
transfer or during a possible phase of reverse mass transfer, and we
assume that the entire envelope of the primary becomes available for star
formation.
%
We assume helium-core masses as in \citet{Prantzos+Charbonnel06} for
stars more massive than 10\Msun. Under these assumptions, the slow
ejecta of massive binaries represent 13\% of the mass originally
present in stars: more than the ejecta of AGB and spin stars combined.


Measurements of lithium suggest that the ejecta of the first generation are diluted with pristine gas \citep{Pasquini+05}.
Together with an equal amount of pristine gas, the ejecta of binaries
with donors more massive than 10\Msun~would represent 26\% of the
initial cluster mass (compared to 38\% contained in the first
generation of low-mass stars). 
The adopted lower mass limit for our binary scenario is rather arbitrary.  If we take the {potential} contribution of intermediate mass stars (4-10\Msun)  into account according to this scenario, there would be enough ejecta to form a second population of chemically peculiar stars that outnumbers the first generation of normal stars. 
 { The assumptions in this scenario can be relaxed if the evaporation of stars from the cluster primarily affects the first stellar generation, as suggested by \citet{D'Ercole+08} and \citet{Decressin+08} .  }

\section{Conclusions}
We propose massive binaries as a source for the internal
pollution of globular clusters. The majority of massive stars are
expected to be members of interacting binary systems. These return
most of the envelope of their primary star to the interstellar medium
during non conservative mass transfer.  We show that there may be more
polluted material ejected by binaries than by the two
previously suggested sources: massive AGB stars and the slow winds of
fast-rotating massive stars. After dilution with pristine material, as
lithium observations suggest, binaries could return enough
material to form a chemically enriched second generation
that is as numerous as the first generation of low-mass stars,
without the need to assume a highly anomalous IMF, external
pollution of the cluster or a significant loss of stars from the
unenriched first generation.

In addition to providing a new source of slowly-ejected enriched
material, binary interaction also affects the previously proposed
scenarios.  Binary mass transfer naturally produces a large number of
fast-rotating massive stars that may enrich their surroundings even
more.  Binary interaction will also affect the yields of
intermediate-mass stars. Premature ejection of the envelope in
4-9\Msun~stars will result in ejecta with less pronounced
anti-correlations, as suggested in the AGB scenario. On the other hand,
we expect that binary-induced mass loss may also prevent the dredge-up
of helium-burning products. 

For a detailed comparison of the chemical predictions of this scenario, binary models {for a range of masses and orbital periods} are needed and population synthesis models are essential to fullly evaluate the mass budget of the different sources.  {Finally, some peculiar feature, such as the apparent presence of distinct, chemically homogeneous subpopulations in  $\omega$~Cen and NGC~2808 \citep[e.g.][]{Renzini08} deserves further attention. }

\begin{acknowledgements}
We thank T. Decressin,  E.~Glebbeek,  A.~Karakas,  C. Charbonnel,  B.~van Veelen, M.~Cantiello, {and the referee  F.~D'Antona} for useful discussions.
\end{acknowledgements}

\bibliographystyle{aa}
\bibliography{references}
\end{document}